\documentclass{ifacconf}

\usepackage{natbib}        % required for bibliography
\usepackage{amssymb}
\usepackage{amsmath}
\usepackage{amsmath,bm}
\usepackage{braket}
\usepackage{mathrsfs}
\begin{document}
\begin{frontmatter}

\title{Complete Positivity Violation \\ in Higher-order \\ Quantum Adiabatic Elimination\thanksref{footnoteinfo}}
% 8 words

\thanks[footnoteinfo]{  This work has been supported by the Engineering for Quantum Information Processors (EQIP) Inria challenge project, the European Research Council (ERC) under the European Union's Horizon 2020 research and innovation programme (grant agreement No. [884762]), and French Research Agency through the ANR grant HAMROQS.}

\author[Paris]{Masaaki Tokieda}
\author[Lyon]{Cyril Elouard}
\author[Paris,Ghent]{Alain Sarlette}
\author[Paris]{Pierre Rouchon}

\address[Paris]{Laboratoire de Physique de l'École Normale Supérieure, Inria, ENS, Mines Paris, PSL, CNRS, Université PSL, Paris, France}
\address[Lyon]{Inria, ENS Lyon, LIP, F-69342, Lyon Cedex 07, France}
\address[Ghent]{Department of Electronics and Information Systems, Ghent University, Belgium}

\begin{abstract}                % 50 - 100 words
  When a composite Lindblad system consists of weakly coupled sub-systems with fast and slow timescales, the description of slow dynamics can be simplified by discarding fast degrees of freedom. This model reduction technique is called adiabatic elimination.
  While second-order perturbative expansion with respect to the timescale separation has revealed that the evolution of a reduced state is completely positive, this paper presents an example exhibiting complete positivity violation in the fourth-order expansion.
  Despite the non-uniqueness of slow dynamics parametrization, we prove that complete positivity cannot be ensured in any parametrization.
  The violation stems from correlation in the initial state.
\end{abstract}  % 99 words


\end{frontmatter}
%===============================================================================

\section{Introduction}

Quantum systems interacting with a surrounding environment are called open quantum systems.
Their studies have attracted considerable attention not only because perfect isolation of a quantum system is unrealistic, but also because control and measurement of a target quantum state are mostly achieved by coupling a target system to another one.
States of an open quantum system are described by a density matrix, which is Hermitian, unit-trace, and positive semidefinite.
Any operation describing a physical process should thus preserve those properties.
The condition of positivity is usually replaced by complete positivity (\cite{NielsenChuang}).
Complete positivity is a property of a linear map on matrix spaces.
Let $\Phi$ be a linear map $\Phi: \mathbb{C}^{n \times n} \to \mathbb{C}^{m \times m}$ and $\mathcal{I}_n$ be the identity map on $\mathbb{C}^{n \times n}$.
$\Phi$ is positive when $\Phi(A)$ is positive semidefinite for all positive semidefinite $A$, and $\Phi$ is completely positive when $\Phi \otimes \mathcal{I}_p: \mathbb{C}^{n \times n} \otimes \mathbb{C}^{p \times p} \to \mathbb{C}^{m \times m} \otimes \mathbb{C}^{p \times p}$ with the tensor product $\otimes$ is positive for any positive integer $p$.
The complete positivity requirement in physics stems from the fact that a density matrix of an open quantum state is a reduced one, and the total density matrix including an environment should also be positive semidefinite.
The complete positivity condition is imposed in deriving a Linbdlad equation, which has widely been used to simulate the evolution of open quantum systems.
One can show that the evolution of a density matrix is governed by a Lindblad equation if and only if the time evolution map is a trace preserving completely positive map, also called a Kraus map, and satisfies the semi-group relation at any time (\cite{GKS}, \cite{Havel03}, \cite{Lindblad}).

In this paper, we consider a composite open quantum system where the total evolution is governed by a Lindblad equation.
The composite system is assumed to consist of a fast decaying sub-system being weakly coupled to another system with a slower time scale.
In this setting, the time evolution typically starts with decay of fast degrees of freedom followed by a slower time evolution of the remaining slow degrees of freedom.
In capturing the slow dynamics, thus, the fast degrees of freedom can be discarded.
This model reduction technique is known in quantum physics as adiabatic elimination.
One of its important applications is reservoir engineering, in which the dissipative dynamics of the slow system is designed by crafting the coupling.
This idea has been used, for instance, to confine the state of a quantum harmonic oscillator to the cat qubit manifold (\cite{Mazyar14}).

For general settings, adiabatic elimination was formulated using degenerate perturbation theory in (\cite{Zanardi16}).
Later, \cite{Azouit17} provided a geometric picture based on center manifold theory (\cite{Fenichel79}).
The system according to this theory does feature (exactly) an invariant manifold (in fact, subspace) corresponding to slow dynamics, and hence we can view the model reduction to slow degrees of freedom as trying to approximate this manifold and the evolution once we would initialize the system on it.
To formulate the model reduction based on this picture, we parametrize the degrees of freedom on an invariant manifold.
We then seek to find two maps; one describing the time evolution of the parameters and the other assigning the parametrization to the solution of the Lindblad equation, that is, the density matrix of the total system.
To compute these maps approximately for general problems, the asymptotic expansion with respect to the timescale separation is performed.
In this way, \cite{Azouit17} established a way to compute higher-order contributions systematically.

In the geometric approach, adiabatic elimination includes a gauge degree of freedom associated with the non-uniqueness of the parametrization.
If the slow dynamics is parametrized via a density matrix, then one expects as a physical requirement that the two maps introduced above should preserve the quantum structure.
This expectation is behind the conjecture made in \cite{Azouit17}; the authors conjectured the existence of a gauge choice such that the reduced dynamics is governed by a Lindblad equation and the assignment is a Kraus map up to any order of the asymptotic expansion.
So far, this has been proved to be true for general settings up to the second-order expansion;
it was shown in \cite{Azouit17} that the evolution equation admits a Lindblad equation and in \cite{AzouitThesis} that there always exists a gauge choice ensuring the Kraus map assignment.
For a two qubit system, \cite{Alain20} reported an example supporting the conjecture at any order.

In this paper, however, we refute this conjecture by showing a counterexample.
To be more precise, we prove that, with fourth-order contributions, complete positivity of the reduced dynamics cannot be ensured, whatever the gauge choice is.
We start by showing that gauge transformation does not change the spectrum of the time evolution map, which can be directly computed in a certain gauge choice.
Then, the conjecture is disproved if we can show that there does not exist a Kraus map which has the same spectrum as the time evolution map.
For qubit systems, such inverse eigenvalue problem was solved in \cite{WPG10}, where the authors revealed necessary and sufficient conditions for the existence of a Kraus map.
To exploit their result, we consider an oscillator-qubit system in which the qubit system is the slow one.
In a gauge choice, this system leads to reduced dynamics described by a non-Lindblad equation.
We then prove the impossibility of ensuring a complete positive evolution in any gauge choice.
We discuss this complete positivity violation in terms of the correlation between the fast and slow degrees of freedom, which imposes a restriction on the initial state of the reduced system.

The reminder of this paper is organized as follows.
Sec.\ref{sec:AE} reviews the machinery of adiabatic elimination developed in \cite{Azouit17} and shows how gauge transformation affects the two maps.
Sec.\ref{sec:JC} introduces a qubit-oscillator system to be investigated.
Sec.\ref{sec:result} presents our main result.
In Sec.\ref{sec:diss}, we discuss interpretation of the results.

\section{Adiabatic Elimination}
\label{sec:AE}

Let $\mathscr{H}_A$ ($\mathscr{H}_B$) be the Hilbert space of the fast (slow) sub-system.
The density matrix of the composite system, $\mathscr{H}_A \otimes \mathscr{H}_B$, denoted by $\rho$ follows a Lindblad equation,
\begin{equation}
  \frac{d}{dt} \rho = \mathcal{L}_A \otimes \mathcal{I}_B (\rho) + \epsilon \mathcal{I}_A \otimes \mathcal{L}_B (\rho) + \epsilon \mathcal{L}_{\rm int} (\rho) \equiv \mathcal{L}_{\rm tot} (\rho).
  \label{eq:AE_master}
\end{equation}
For $\xi = A$ and $B$, $\mathcal{I}_\xi$ are the identity superoperators acting only to operators on $\mathscr{H}_\xi$.
$\mathcal{L}_A$ is a Lindbladian acting only on $\mathscr{H}_A$ and generally reads
\begin{equation}
  \mathcal{L}_A = - i H_A^\times + \sum_{k} \mathcal{D}[L_{A,k}],
  \label{eq:AE_LA}
\end{equation}
with a Hamiltonian $H_A$ and jump operators $\{ L_{A,k} \}$, all of which are operators on $\mathscr{H}_A$.
We have also introduced the commutator superoperator $H^\times (\bullet) = H \bullet - \bullet H$ and the dissipator superoperator $\mathcal{D}[L] (\bullet) = L \bullet L^\dagger - (L^\dagger L \bullet + \bullet L^\dagger L)/2$.
We assume that the evolution only with $\mathcal{L}_A$ exponentially converges to a unique steady state $\bar{\rho}_A$.
In other words, among the spectrum of $\mathcal{L}_A$, the eigenvalue zero is simple and the other eigenvalues have strictly negative real part.
$\mathcal{L}_B$ and $\mathcal{L}_{\rm int}$ are superoperators acting on $\mathscr{H}_B$ and $\mathscr{H}_A \otimes \mathscr{H}_B$, respectively, and are assumed to contain only Hamiltonian terms.
$\epsilon$ is a non-negative parameter representing the timescale separation.
Physically, $\mathcal{L}_A$ and $\mathcal{L}_B$ describe the internal dynamics of $\mathscr{H}_A$ and $\mathscr{H}_B$, respectively, and $\mathcal{L}_{\rm int}$ determines how the two sub-systems interact.

As described in the Introduction section, we use a density matrix to parametrize the degrees of freedom on an invariant manifold.
Let us denote the parametrization by $\rho_s$ in general.
One possible choice, which has commonly been used to represent the reduced dynamics, is the partial trace $\rho_s = {\rm tr}_A (\rho)$ with ${\rm tr}_A$ the trace over $\mathscr{H}_A$.
This choice plays a central role in the following discussions.
For clear distinction, we denote the partial trace by $\rho_B = {\rm tr}_A (\rho)$ and general parametrization by $\rho_s$.

Once the parametrization is fixed to $\rho_s$, we seek to find the following two maps.
One, denoted by $\mathcal{L}_s$, describes the time evolution of $\rho_s$, namely, $(d/dt) \rho_s = \mathcal{L}_s (\rho_s)$.
The other, denoted by $\mathcal{K}$, assigns $\rho_s$ to the solution of the Lindblad equation $\rho$, $\rho = \mathcal{K} (\rho_s)$.
Throughout this paper, we assume that $\mathcal{K}$ and $\mathcal{L}_s$ are linear and time-independent (see below).
Since $\rho$ satisfies (\ref{eq:AE_master}), we obtain
\begin{equation}
  \mathcal{K} (\mathcal{L}_s (\rho_s)) = \mathcal{L}_{\rm tot} (\mathcal{K}(\rho_s)).
  \label{eq:AE_inv1}
\end{equation}
To compute $\mathcal{K}$ and $\mathcal{L}_s$ satisfying this relation approximately, we assume $\epsilon \ll 1$ and perform the asymptotic expansions as
\begin{equation}
  \mathcal{K} = \sum_{n = 0}^\infty \epsilon^n \mathcal{K}_n, \ \ \ \ \ \ \ \ \mathcal{L}_s = \sum_{n = 0}^\infty \epsilon^n \mathcal{L}_{s,n}.
  \label{eq:AE_asymexp}
\end{equation}
When $\epsilon = 0$, the solution of (\ref{eq:AE_master}) after the decay of the fast sub-system reads $\bar{\rho}_A \otimes {\rm tr}_A (\rho (t=0))$, with the initial density matrix $\rho(t=0)$.
Therefore, the zeroth elements are given by
\begin{equation}
  \mathcal{K}_{0} (\rho_s) = \bar{\rho}_A \otimes \rho_s, \ \ \ \ \ \ \ \ \mathcal{L}_{s,0} (\rho_s) = 0.
  \label{eq:AE_asymexp0}
\end{equation}

Higher-order contributions can be computed by substituting the expansions (\ref{eq:AE_asymexp}) into (\ref{eq:AE_inv1}).
Here, we particularly focus on the computation of $\mathcal{K}_n$ in order to introduce the gauge degree of freedom.
The $n$-th order of $\epsilon$ reads
\begin{equation}
  \mathcal{L}_A \otimes \mathcal{I}_B (\mathcal{K}_{n} (\rho_s)) =  \bar{\rho}_A \otimes \mathcal{L}_{s,n} (\rho_s) - \mathcal{L}_n(\rho_s),
  \label{eq:AE_LAKn}
\end{equation}
with
\begin{equation*}
  \mathcal{L}_n (\rho_s) = (\mathcal{I}_A \otimes \mathcal{L}_B + \mathcal{L}_{\rm int}) (\mathcal{K}_{n-1}(\rho_s)) - \sum_{k=1}^{n} \mathcal{K}_{k} (\mathcal{L}_{s,n-k}(\rho_s)).
\end{equation*}
To solve the linear equation (\ref{eq:AE_LAKn}) for $\mathcal{K}_{n} (\rho_s)$, one needs to invert $\mathcal{L}_A$.
Note that $\mathcal{L}_A$ is singular since one of the eigenvalue is zero.
Thus, this linear equation is underdetermined.
As shown in \cite{Azouit17}, $\mathcal{K}_{n} (\rho_s)$ is determined only up to $\bar{\rho}_A \otimes {\rm tr}_A (\mathcal{K}_{n} (\rho_s))$ by solving (\ref{eq:AE_LAKn}).
We introduce the undetermined part, $G_n = {\rm tr}_A \circ \mathcal{K}_{n}$, which can be any superoperator on $\mathscr{H}_B$.
This gauge degree of freedom is associated with the non-uniqueness of the parametrization.

From (\ref{eq:AE_LAKn}), one can compute $\mathcal{K}_n$ and $\mathcal{L}_{s,n}$ up to a desired order.
By writing down their forms explicitly, one finds that $\mathcal{K}_n$ and $\mathcal{L}_{s,n+1}$ depend on $G_1, \dots, G_n$.
Therefore, $\mathcal{K}_n$ and $\mathcal{L}_{s,n+1}$ become, for instance, nonlinear functions of $\rho_s$ if so is one of $G_1, \dots, G_n$.
To meet the conditions that $\mathcal{K}$ and $\mathcal{L}_s$ are linear and time-independent, we assume that $\{ G_n \}$ have those properties.

From the definition of $G_n$, we find
\begin{equation}
  \rho_B = {\rm tr}_A (\rho) = {\rm tr}_A (\mathcal{K} (\rho_s)) = \rho_s + G(\rho_s)
  \label{eq:AE_rhobrhos}
\end{equation}
with $G = \sum_{n=1}^\infty \epsilon^n G_n$.
Notice that this definition of the gauge superoperator is different from that in \cite{Azouit17}.
In this paper, $G = 0$ corresponds to the parametrization via the partial trace $\rho_s = \rho_B$.
We further find the following relation about the gauge dependence;

\begin{lem}
  Let us denote the gauge dependence of $\mathcal{K}$ and $\mathcal{L}_s$ explicitly as $\mathcal{K}^G$ and $\mathcal{L}_{s}^{G}$.
  Then, we have
  \begin{equation}
    \mathcal{K}^G = \mathcal{K}^{G=0} \circ (\mathcal{I}_B + G),
    \label{eq:AE_KG}
  \end{equation}
  and, when $(\mathcal{I}_B + G)$ is invertible,
  \begin{equation}
    \mathcal{L}_{s}^G = (\mathcal{I}_B + G)^{-1} \circ \mathcal{L}_{s}^{G=0} \circ (\mathcal{I}_B + G).
    \label{eq:AE_LG}
  \end{equation}
  \label{theorem:gaugedep}
\end{lem}
\begin{pf}
  For $\mathcal{K}^G$, note $\rho = \mathcal{K}^{G} (\rho_s) = \mathcal{K}^{G=0} (\rho_B)$.
  Substituting (\ref{eq:AE_rhobrhos}) into the rightmost side gives (\ref{eq:AE_KG}).
  For $\mathcal{L}_{s}^G$, we use $\mathcal{L}_{s}^{G=0} (\rho_B) = (d/dt) \rho_B = (d/dt) (\rho_s + G(\rho_s)) = \mathcal{L}_{s}^{G} (\rho_s) + G (\mathcal{L}_{s}^{G} (\rho_s))$.
  Comparing the leftmost and rightmost sides gives $\mathcal{L}_{s}^{G=0} \circ (\mathcal{I}_B + G) = (\mathcal{I}_B + G) \circ \mathcal{L}_{s}^{G}$.
  Thus, when $(\mathcal{I}_B + G)$ is invertible, we obtain (\ref{eq:AE_LG}).
  $ \ $ Q.E.D.
\end{pf}

We make three remarks about these results.
First, these relations are results of general basis change and have nothing to do with the quantum structure.
Second, from $G = \sum_{n=1}^\infty \epsilon^n G_n$, the existence of $(\mathcal{I}_B + G)^{-1}$ is guaranteed as long as $\epsilon \ll 1$ such that the truncation at a finite order is reasonable.
Third, (\ref{eq:AE_LG}) means that the spectrum of $\mathcal{L}_s$ is independent of gauge choice.
This is expected since decay rate on an invariant manifold must be independent of the way its degrees of freedom are parametrized.

As summarized in the Introduction section, \cite{Azouit17} conjectured the existence of a gauge choice leading to reduced dynamics described by a Lindbladian, $\sum_{j=1}^n \mathcal{L}_{s,j} (\rho_s) = - i H_s^\times (\rho_s) + \sum_{k} \mathcal{D}[L_{s,k}] (\rho_s)$ with a Hamiltonian $H_s$ and jump operators $\{ L_{s,k} \}$,
and assignment described by a Kraus map, $\sum_{j=1}^n \mathcal{K}_{j} (\rho_s) = \sum_{k} M_k \rho_s M_k^\dagger$ with operators $M_k: \mathscr{H}_B \to \mathscr{H}_A \otimes \mathscr{H}_B$, up to $\epsilon^n$ for any positive integer $n$.
For general settings, this conjecture has been proved up to $n = 2$ so far.
In the following sections, we present an example where this conjecture does not hold true with up to $n = 4$ terms.

\section{Problem setting}
\label{sec:JC}

We consider an oscillator-qubit system in which the dissipative oscillator system is eliminated.
The Hamiltonian is given by the Jaynes-Cummings Hamiltonian and the oscillator is coupled to a Markovian environment at finite temperature.
We assume that the qubit is non-dissipative for simplicity.
In the frame rotating with the qubit frequency, we have
\begin{equation*}
  \begin{array}{cc}
    \mathcal{L}_{A} = - i\Delta_A (a^\dagger \! a)^\times + \gamma (1 + n_{\rm th}) \mathcal{D}[a] + \gamma n_{\rm th} \mathcal{D}[a^\dagger],
  \end{array}
  \label{eq:JC_LA}
\end{equation*}
\begin{equation*}
  \epsilon \, \mathcal{L}_{\rm int} = - i g (a^\dagger \otimes \sigma_{-} + a \otimes \sigma_{+})^\times,
\end{equation*}
and $\mathcal{L}_B = 0$, with the oscillator detuning from the qubit frequency $\Delta_A$, the decay rate $\gamma$, the asymptotic oscillator number in the absence of coupling $n_{\rm th}$, and the coupling strength $g$.
$a$ and $a^\dagger$ are the annihilation and creation operators of the oscillator, respectively, and $\sigma_{\pm} = (\sigma_x \pm i \sigma_y)/2$ with the Pauli matrices $\{ \sigma_i \}_{i=x,y,z}$.
This form of Lindbladian is used as a benchmark when analyzing oscillator-qubit interacting systems in cavity or superconducting circuit architecture.

The full spectrum and eigenoperators of $\mathcal{L}_A$ can be obtained with various methods, such as the third quantization (\cite{Prosen10}).
Result confirms, as long as $\gamma > 0$, the existence of a unique steady state $\bar{\rho}_A$ given by
\begin{equation*}
  \bar{\rho}_A =  \Big( \frac{n_{\rm th}}{1+n_{\rm th}} \Big)^{a^\dagger \! a} / {\rm tr}_A \Big[ \Big( \frac{n_{\rm th}}{1+n_{\rm th}} \Big)^{a^\dagger \! a} \Big].
\end{equation*}

\section{Results of fourth-order adiabatic elimination}
\label{sec:result}

The timescale of the oscillator system is characterized by $\gamma^{-1}$, while that of the interaction is $|g|^{-1}$.
Thus, the timescale separation parameter $\epsilon$ reads $\epsilon = |g|/\gamma$.
Assuming $\epsilon \ll 1$, we compute up to the fourth-order contributions.
As a result, we obtain the following;

\begin{prop}
  Up to the fourth-order expansion, $\mathcal{L}_{s}$ for the partial trace, $\mathcal{L}_{s}^{G=0}$, reads
  \begin{equation}
    \begin{array}{cc}
      \mathcal{L}_{s}^{G=0} = - i \frac{\omega_B^{(4)}}{2} \sigma_z^\times \vspace{0.1em} \\
      + \gamma_{-}^{(4)} \mathcal{D}[\sigma_-] + \gamma_{+}^{(4)} \mathcal{D}[\sigma_+] + \gamma_\phi^{(4)} \mathcal{D}[\sigma_z].
    \end{array}
    \label{eq:JC_LsPT}
  \end{equation}
  The coefficients $\omega_B^{(4)}$, $\gamma_{\pm}^{(4)}$, and $\gamma_{\phi}^{(4)}$ are real numbers defined by
  \begin{equation*}
    \omega_B^{(4)} = {\rm Im} (b_- + b_+), \ \ \ \ \ \gamma_{\pm}^{(4)} = 2 {\rm Re} (b_{\pm}),
  \end{equation*}
  and
  \begin{equation*}
    \gamma_\phi^{(4)} = - \frac{8g^4 n_+ n_- ( 3 - 6 (2\Delta_A/\gamma)^2 - (2\Delta_A/\gamma)^4 )}{\gamma^3 (1+(2\Delta_A/\gamma)^2)^3},
  \end{equation*}
  where ${\rm Re}$ (${\rm Im}$) is real (imaginary) part and $b_{\pm}$ are
  \begin{equation*}
    b_\pm =  \frac{2g^2 n_{\pm}}{\bar{\gamma}} + \frac{8g^4 n_{\pm}^2}{\bar{\gamma}^3} + \frac{8g^4 n_+ n_-(1+8 i \gamma \Delta_A / |\bar{\gamma}|^2)}{\bar{\gamma}^* |\bar{\gamma}|^2},
  \end{equation*}
  with $n_+ = n_{\rm th}$, $n_- = 1 + n_{\rm th}$, and $\bar{\gamma} = \gamma + 2i\Delta_A$.
  \label{theorem:4thorderL}
\end{prop}
\begin{pf}
  From the spectral decomposition, we can evaluate the Moore-Penrose inverse of $\mathcal{L}_A$, with which the linear equation (\ref{eq:AE_LAKn})  can be solved for $\mathcal{K}_n$ under the condition ${\rm tr}_A \circ \mathcal{K}_n = 0 \ (n > 0)$.
  Repeating the computation up to $\mathcal{K}_3$ then gives (\ref{eq:JC_LsPT}).
  $ \ $ Q.E.D.
\end{pf}

The $\gamma_{\pm}^{(4)}$ and $\gamma_\phi^{(4)}$ terms describe the effective qubit decay induced by the coupling to the dissipative oscillator.
On one hand, for small enough $|g|/\gamma$ , $\gamma_{\pm}^{(4)}$ are dominated by the second-order contributions given by $\gamma_{\pm}^{(2)} = 4 g^2 \gamma n_{\pm}/|\bar{\gamma}|^2 > 0$, and thus $\gamma_{\pm}^{(4)} > 0$.
On the other hand, $\gamma_\phi^{(4)}$ involves only the fourth-order contribution and
\begin{equation*}
  \begin{array}{cc}
    \gamma_\phi^{(4)} < 0 \ \ \ {\rm when} \\
    n_{\rm th} > 0 \ \ {\rm and} \ \ |\Delta_A|/\gamma < \sqrt{2\sqrt{3}-3}/2 \simeq 0.34,
  \end{array}
  \label{eq:JC_negative}
\end{equation*}
even if the condition for the asymptotic expansion, $|g|/\gamma \ll 1$, holds.

Such negative coefficient does not appear in the second-order expansion.
Before investigating how this affects complete positivity of the reduced dynamics, let us prove stability and positivity of the time evolution map;

\begin{cor}
  For (\ref{eq:JC_LsPT}), the spectrum of the time evolution map, $\exp(\mathcal{L}_{s}^{G=0} t)$ with $t \in \mathbb{R}_{\geq 0}$, is given by
  \begin{equation*}
    \{ 1, e^{-t/T_2 + i \omega_B^{(4)}t}, e^{-t/T_2 - i \omega_B^{(4)} t}, e^{-t/T_1} \},
  \end{equation*}
  with $1/T_1 = \gamma_-^{(4)} + \gamma_+^{(4)}$ and $1/T_2 = 1/(2 T_1) + 2 \gamma_\phi^{(4)}$.
  \label{theorem:TEspectrum}
\end{cor}
\begin{pf}
  With $I_B$ the identity matrix on $\mathscr{H}_B$ ($2$-dimensional identity matrix), the set $\{ I_B/\sqrt{2}, \sigma_x/\sqrt{2}, \sigma_y/\sqrt{2}, \sigma_z/\sqrt{2} \}$ is an orthonormal basis with the Hilbert-Schmidt inner product.
  Let $[\mathcal{L}_{s}^{G=0}]$ be the $4 \times 4$ matrix representation of $\mathcal{L}_{s}^{G=0}$ in this basis. It reads
  \begin{equation*}
    [\mathcal{L}_{s}^{G=0}] =
    \begin{pmatrix}
      0 & 0 & 0 & 0 \\
      0 & -1/T_2 & - \omega_B^{(4)} & 0 \\
      0 & \omega_B^{(4)} & - 1/T_2 & 0 \\
      R_{z}/T_1 & 0 & 0 & - 1/T_1
    \end{pmatrix},
  \end{equation*}
  with $R_{z} = - (\gamma_-^{(4)} - \gamma_+^{(4)}) T_1$, the meaning of which will become clear below.
  From this, the spectrum of $\mathcal{L}_{s}^{G=0}$ is given by $\{ 0, -1/T_2 + i \omega_B^{(4)}, -1/T_2 - i \omega_B^{(4)}, -1/T_1 \}$.
  By multiplying $t$ and then exponentiating them, we obtain the above result.
  $ \ $ Q.E.D.
\end{pf}

Since $\gamma_{\pm}^{(4)} > 0$ and $\gamma_{\pm}^{(4)} \gg |\gamma_\phi^{(4)}|$ in the parameter region of interest, we have $T_1 > 0$ and $T_2 > 0$.
Therefore, the time evolution is stable even when $\gamma_\phi^{(4)}$ is negative.
In addition, we can show that it preserves positivity;

\begin{cor}
  For (\ref{eq:JC_LsPT}), the time evolution map $\exp(\mathcal{L}_{s}^{G=0} t)$ is positive for any $t \in \mathbb{R}_{ \geq 0}$.
  \label{theorem:positive}
\end{cor}
\begin{pf}
  Let $\bm{r}(t) = (r_x(t), r_y(t), r_z(t))^\top \in \mathbb{R}^3$ with the matrix transpose $\top$ be the Bloch vector , which is related to the partial trace as $\rho_B (t) = (I_B + \sum_{i = x,y,z} r_i(t) \sigma_i)/2$.
  From (\ref{eq:JC_LsPT}), the evolution of the Bloch vector reads
  \begin{equation*}
    \frac{d}{dt} r_x(t) = - \frac{r_x(t)}{T_2} - \omega_B^{(4)} r_y(t), \frac{d}{dt} r_y(t) = - \frac{r_y(t)}{T_2} + \omega_B^{(4)} r_x(t),
  \end{equation*}
  and
  \begin{equation*}
    \frac{d}{dt} r_z(t) = - \frac{1}{T_1} (r_z(t) - R_{z}).
  \end{equation*}
  From these equations, we find that $R_{z}$ is the asymptotic value of $r_z(t)$, $\lim_{t \to \infty} \bm{r}(t) = (0, 0, R_{z})^\top$.
  They also lead to
  \begin{equation*}
    \frac{d}{dt} \bm{r}^2(t) = - 2 \Big( \frac{\bm{r}^2(t) - r_z^2(t)}{T_2} + \frac{r_z(t) (r_z(t) - R_{z})}{T_1} \Big).
  \end{equation*}
  The partial trace $\rho_B(t)$ is positive semidefinite if and only if $\bm{r}^2(t) \leq 1$.
  If $(d/dt) \bm{r}^2(t) < 0$ with the constraint $\bm{r}^2(t) = 1$ for any $t$, then we have $\bm{r}^2(t) \leq 1$ and thus the time evolution map is positive.
  Substituting $\bm{r}^2(t) = 1$ into the above equation, we obtain
  \begin{equation*}
    \begin{array}{cc}
      \frac{d}{dt} \bm{r}^2(t) = \vspace{0.2em} \\
      - \frac{2}{\Delta T}  \Big[ \Big( r_z(t) - \frac{\Delta T}{2 T_1} R_{z} \Big)^2 + (\Delta T)^2 ( \gamma_-^{(4)} \gamma_+^{(4)} - 4 (\gamma_\phi^{(4)})^2 ) \Big],
    \end{array}
  \end{equation*}
  with $1/\Delta T = 1/T_1 - 1/T_2 = (\gamma_-^{(4)} + \gamma_+^{(4)})/2 - 2 \gamma_\phi^{(4)}$.
  In the parameter region of interest, we have $\gamma_{\pm}^{(4)} \gg |\gamma_\phi^{(4)}|$.
  This leads to $\Delta T > 0$ and $\gamma_-^{(4)} \gamma_+^{(4)} > 4 (\gamma_\phi^{(4)})^2$.
  Therefore, $(d/dt) \bm{r}^2(t) < 0$ whenever $\bm{r}^2(t) = 1$.
  $ \ $ Q.E.D.
\end{pf}

Now we investigate complete positivity.
To judge whether (\ref{eq:JC_LsPT}) is in the Lindblad form or not, the following lemma, which follows from \cite{Havel03}, is sufficient;
\begin{lem}
  Let $\mathscr{H}$ be a Hilbert space and $\mathcal{L}$ the superoperator on $\mathscr{H}$ whose operation is given by
  \begin{equation*}
    \mathcal{L}(\bullet) = \sum_{\alpha,\beta = 1}^{D} \Gamma_{\alpha,\beta} \Big[ A_\alpha \bullet A_\beta^\dagger - \frac{1}{2} ( A_\beta^\dagger A_\alpha \bullet + \bullet A_\beta^\dagger A_\alpha ) \Big],
  \end{equation*}
  with $\{ A_\alpha \}$ being operators on $\mathscr{H}$, a positive integer $D$, and a $D$-dimensional Hermitian matrix $\Gamma$.
  Suppose that $\{ A_\alpha \}$ are traceless and linearly independent.
  Then, $\mathcal{L}$ is in the Lindblad form if and only if $\Gamma$ is positive semidefinite.
  \label{theorem:LindbladForm}
\end{lem}

Care should be taken when $\{ A_\alpha \}$ are linearly dependent or have non-zero trace.
In that case, $\Gamma$ being positive semidefinite is only a sufficient condition for $\mathcal{L}$ to be a Lindbladian, but not necessary in general.

\begin{thm}
  If $\gamma_\phi^{(4)} < 0$ in (\ref{eq:JC_LsPT}), then $\mathcal{L}_{s}^{G=0}$ is not in the Lindblad form.
  \label{theorem:non-Lindblad}
\end{thm}
\begin{pf}
  From Lemma \ref{theorem:LindbladForm} and the fact that $\sigma_{\pm}$ and $\sigma_z$ are traceless and linearly independent.
  $ \ $ Q.E.D.
\end{pf}

To be more precise, it is complete positivity of the evolution that is violated;

\begin{cor}
  If $\gamma_\phi^{(4)} < 0$ in (\ref{eq:JC_LsPT}), then the time evolution map $\exp(\mathcal{L}_{s}^{G=0} t)$ is not completely positive at non-zero but infinitesimal time $t$.
\end{cor}
\begin{pf}
  As shown in \cite{GKS}, \cite{Havel03}, and \cite{Lindblad}, the generator is in the Lindblad form if and only if the time evolution map is a Kraus map and satisfies the semigroup relation at any time.
  The time evolution map $\exp(\mathcal{L}_{s}^{G=0} t)$ satisfies the semigroup relation and preserves the Hermitian property and trace.
  Thus, Theorem \ref{theorem:non-Lindblad} signifies the violation of complete positivity at some time.
  A finite time evolution can be obtained by concatenating infinitesimal time evolution owing to the semigroup relation.
  Since concatenation of completely positive maps is completely positive (\cite{Havel03}), the violation occurs at infinitesimal time.
  $ \ $ Q.E.D.
\end{pf}

When $\gamma_{\phi}^{(4)}$ is negative, the time evolution map $\exp(\mathcal{L}_{s}^{G=0} t)$ at infinitesimal $t$ is positive, while it is not completely positive.
It is widely recognized that the matrix transpose has such property (\cite{NielsenChuang}).
Here, we have found another example which is derived naturally from a physics equation.

So far, we have considered the evolution of the partial trace.
Strikingly, the above result of the non-Lindblad form can be generalized to any gauge choice.
The proof proceeds as we have summarized at the end of the Introduction section.
We first recall the following result in \cite{WPG10};

\begin{prop}
  (\cite{WPG10}) Given $\Lambda \in \mathbb{C}^4$, the following statements are equivalent:
  \begin{itemize}
    \item There exists a Kraus map the spectrum of which is given by $\Lambda$.
    \item $\Lambda = {1} \cup \lambda$ where $\lambda \in \mathbb{C}^3$ is closed under complex conjugation.
    Furthermore, if we define $s \in \mathbb{R}^3$ by $s_i = \lambda_i$ if $\lambda_i \in \mathbb{R}$ and $s_i = |\lambda_i|$ otherwise, then
    \begin{equation}
      s \in \mathcal{T},
      \label{eq:JC_thm}
    \end{equation}
    where $\mathcal{T} \subset \mathbb{R}^3$ is the tetrahedron whose corners are $(1,1,1)$, $(1,-1,-1)$, $(-1,1,-1)$, and $(-1,-1,1)$.
  \end{itemize}
  \label{theorem:WPG10}
\end{prop}

With this, we come to the main result of this paper;

\begin{thm}
  If $\gamma_\phi^{(4)} < 0$ in (\ref{eq:JC_LsPT}), then $\mathcal{L}_{s}^{G}$ is not in the Lindblad form in any gauge choice $G$.
  \label{theorem:nonLindbladG}
\end{thm}
\begin{pf}
  We use Proposition \ref{theorem:WPG10}.
  Let $\Lambda$ be the spectrum of the time evolution map $\exp( \mathcal{L}_{s}^{G=0} t )$. From Corollary \ref{theorem:TEspectrum},
  \begin{equation*}
    \Lambda = \{ 1, e^{-t/T_2 + i \omega_B^{(4)}t}, e^{-t/T_2 - i \omega_B^{(4)} t}, e^{-t/T_1} \},
  \end{equation*}
  which is closed under complex conjugation.
  From the definition, $\{ s_i \}$ are now all positive.
  In this case, assuming $s_1 \geq s_2 \geq s_3$, the condition (\ref{eq:JC_thm}) reads $s_1 \leq 1$ and $s_1 + s_2 \leq 1 + s_3$ (see (\cite{WPG10})).
  When $\gamma_\phi^{(4)} < 0$, we always have $1/T_1 > 1/T_2$, and $\exp( -t/T_2) > \exp( -t/T_1 )$ for $t > 0$.
  Thus, we set $s_1 = s_2 = \exp( -t/T_2)$ and $s_3 = \exp( -t/T_1)$.
  Then, while the first condition $s_1 \leq 1$ is satisfied, the second condition reads
  \begin{equation}
    2 e^{-t/T_2} \leq 1 + e^{-t/T_1}.
    \label{eq:JC_minTcp}
  \end{equation}
  For infinitesimal $t$ such that $\exp( -t/T_i ) \simeq 1 - t/T_i \ (i = 1,2)$,
  this condition reads $\gamma_\phi^{(4)} \geq 0$ and is violated if $\gamma_\phi^{(4)} < 0$.
  Therefore, there does not exist a Kraus map which has the same spectrum as $\exp( \mathcal{L}_{s}^{G=0} t)$ at infinitesimal time $t$.
  From (\ref{eq:AE_LG}), the gauge transformation induces similarity transformation of the time evolution map,
  \begin{equation*}
    e^{\mathcal{L}_s^G t} = (\mathcal{I}_B + G)^{-1} \circ e^{\mathcal{L}_s^{G = 0} t} \circ (\mathcal{I}_B + G).
  \end{equation*}
  Here, we have assumed that $(\mathcal{I}_B + G)$ is invertible, which is a valid assumption in the parameter region of interest as discussed below Lemma \ref{theorem:gaugedep}.
  This proves that $\exp (\mathcal{L}_s^G t)$ at infinitesimal time $t$ is not a Kraus map in any choice of gauge $G$.
  $ \ $ Q.E.D.
\end{pf}

Therefore, this oscillator-qubit system serves as a counterexample to the conjecture in \cite{Azouit17}.

\section{Discussions}
\label{sec:diss}

In this section, we discuss interpretation of the results found so far.
As we have seen in the proof of Theorem \ref{theorem:nonLindbladG}, complete positivity violation in any gauge choice stems essentially from that in the partial trace parametrization.
Accordingly, we concentrate on the partial trace in this section.

We first note that the complete positivity violation is not an artifact caused by truncating the series expansion at a finite order.
Higher-order contributions only correct the values of $1/T_1$ and $1/T_2$ in (\ref{eq:JC_minTcp}).
Thus, for small enough $|g|/\gamma$, they cannot cure the violation of the inequality at the fourth-order.
In fact, we have found an example where adiabatic elimination can be performed at any order, yet the partial trace evolution is not completely positive (\cite{Tokieda22}).

As shown in Corollary \ref{theorem:positive}, the time evolution map for the oscillator-qubit system is positive even when it is not completely positive.
As long as the time evolution map is positive, the density matrix properties are preserved along the evolution, and result can be interpreted physically.
However, this positivity is not generally the case.
In \cite{Tokieda22}, we present an example where the time evolution map is not even positive.

In order to understand its origin, let us first recall the reason why we expected a completely positive evolution.
It can be seen by writing the time evolution of the partial trace via that of the total state.
If we assume a separable initial state with $\bar{\rho}_A$, we have
\begin{equation}
  \rho_B (t) = {\rm tr}_A \circ e^{\mathcal{L}_{\rm tot} t} (\bar{\rho}_A \otimes \rho_B(t = 0)),
  \label{eq:diss_InitialSeparable}
\end{equation}
where $\rho_B(t = 0)$ is the partial trace of an initial total state $\rho(t = 0)$, $\rho_B(t = 0) = {\rm tr}_A (\rho(t = 0))$.
Note that ${\rm tr}_A$, $\exp( \mathcal{L}_{\rm tot} t)$, and the map that sends $\rho_B(t = 0)$ to $\bar{\rho}_A \otimes \rho_B(t = 0)$ are all completely positive.
Therefore, the time evolution map from an initial state $\rho_B(t = 0)$ to $\rho_B(t)$ is completely positive.

Here, it should be recalled that, in adiabatic elimination, we initialize a state on an invariant manifold.
When $\epsilon = 0$, a set of density matrices on the invariant manifold is characterized by separable states as $\mathcal{K}_0^{G=0}(\rho_B) = \bar{\rho}_A \otimes \rho_B$ (see (\ref{eq:AE_asymexp0})).
In this case, (\ref{eq:diss_InitialSeparable}) describes the slow dynamics on the invariant manifold.
When higher-order contributions are taken into account, on the other hand, (\ref{eq:diss_InitialSeparable}) includes the transient dynamics of the fast degrees of freedom.
Long after decay time of the fast sub-system, say $t \geq t_{\rm inv}$, the total state $\rho(t)$ is approximately on an invariant manifold.
In this situation, the time evolution map computed in adiabatic elimination is the one that sends $\rho_B(t_{\rm inv}) = {\rm tr}_A (\rho (t_{\rm inv}))$ to $\rho_B(t) \ ( t \geq t_{\rm inv})$.
There is no guarantee that this map is completely positive, even though the map that sends $\rho_B(t=0)$ to $\rho_B(t)$ is completely positive.

To account for the initialization on the invariant manifold, (\ref{eq:diss_InitialSeparable}) should be rewritten as
\begin{equation}
  \rho_B (t) = {\rm tr}_A \circ e^{\mathcal{L}_{\rm tot} t} \circ \mathcal{K}^{G=0} (\rho_B(t = 0)).
  \label{eq:diss_rhobt}
\end{equation}
From this equation, $\mathcal{K}^{G=0}$ is not completely positive if the time evolution map is not.
In fact, it was shown in \cite{AzouitThesis} that one can make $\mathcal{K}^{G}$ completely positive up to second-order by a proper gauge choice, but in general $\mathcal{K}^{G=0}$ is not.\footnote{For a  different choice of $G$, the results are consistent as the map ${\rm tr}_A$ in (\ref{eq:diss_rhobt}) would be replaced by a non-completely positive one.}
For the oscillator-qubit system, this can be seen as follows (we assume $\Delta_A = 0$ for simplicity);
\begin{prop}
  For the oscillator-qubit system, when $\Delta_A = 0$, $\mathcal{K}^{G=0}$ up to the second-order expansion reads
  \begin{equation}
    \begin{array}{ccc}
      \mathcal{K}^{G=0} (\rho_B) = V (\bar{\rho}_A \otimes \rho_B) V^\dagger \vspace{0.1em} \\
      - \frac{4 g^2 (1 + n_{\rm th})}{\gamma^2} (I_A \otimes \sigma_-) (\bar{\rho}_A \otimes \rho_B) (I_A \otimes \sigma_-)^\dagger \vspace{0.2em} \\
      - \frac{4 g^2 n_{\rm th}}{\gamma^2} (I_A \otimes \sigma_+) (\bar{\rho}_A \otimes \rho_B) (I_A \otimes \sigma_+)^\dagger,
    \end{array}
    \label{eq:diss_KG0}
  \end{equation}
  with $I_A$ the identity operator on $\mathscr{H}_A$ and $V = I_A \otimes I_B - (2ig/\gamma) (a^\dagger \otimes \sigma_- + a \otimes \sigma_+) - (2 g^2/\gamma^2) (a^\dagger \! a - n_{\rm th} I_A) \otimes I_B$.
\end{prop}

Because of the minus signs in the last two lines, the second-order expansion of $\mathcal{K}^{G=0}$ is not completely positive.
From the construction, $\mathcal{L}^{G=0}_{s,n}$ with a positive integer $n$ depends on $\{ \mathcal{K}^{G=0}_{k} \}_{k = 0,1,\dots,n-1}$.
For the oscillator-qubit system, we have $\mathcal{L}^{G=0}_{s,2n-1} = 0$.
Thus, complete positivity of the time evolution map can be violated from the fourth-order expansion.
This consideration is consistent with the result in the previous section.
It should be noted that the time evolution map can be completely positive even if $\mathcal{K}^{G=0}$ is not.
If $n_{\rm th} = 0$, for instance, $\gamma_\phi^{(4)} = 0$ and the generator $\mathcal{L}_{s}^{G=0}$ is in the Lindblad form up to the fourth-order terms, despite that $\mathcal{K}^{G=0}$ still contains a negative term.

To understand the origin of non-completely positive $\mathcal{K}^{G=0}$, we note that it is not even positive;

\begin{cor}
  If $\rho_B$ is a pure state, then $\mathcal{K}^{G=0} (\rho_B)$ given by (\ref{eq:diss_KG0}) is not positive semidefinite.
  \label{theorem:PureNegative}
\end{cor}
\begin{pf}
  Let $\ket{0} \in \mathscr{H}_A$ be the vacuum state of the oscillator and $\ket{\psi} \in \mathscr{H}_B$ be a state.
  The matrix element of $\mathcal{K}^{G=0}$ with respect to $\ket{0,\psi} = \ket{0} \otimes \ket{\psi}$ reads
  \begin{equation*}
    \begin{array}{cc}
      \braket{0,\psi|\mathcal{K}^{G=0} (\rho_B)|0,\psi} = \braket{0|\bar{\rho}_A|0} \Big[ \Big( 1 + \frac{4g^2 n_{\rm th}}{\gamma^2} \Big) \braket{\psi|\rho_B|\psi} \vspace{0.2em} \\
      - \frac{4 g^2 (1 + n_{\rm th})}{\gamma^2} \, \braket{\psi_+|\rho_B|\psi_+} - \frac{4 g^2 n_{\rm th}^2}{\gamma^2 (1+n_{\rm th})} \, \braket{\psi_-|\rho_B|\psi_-} \Big].
    \end{array}
  \end{equation*}
  with $\ket{\psi_\pm} = \sigma_\pm \ket{\psi}$.
  If $\rho_B$ is a pure state, its kernel is not empty. Suppose $\ket{\psi}$ is in the kernel of $\rho_B$, that is, $\rho_B \ket{\psi} = 0$.
  Since either $\braket{\psi_+|\rho_B|\psi_+}$ or $\braket{\psi_-|\rho_B|\psi_-}$ is non-zero, we obtain $\braket{0,\psi|\mathcal{K}^{G=0} (\rho_B)|0,\psi} < 0$.
  $ \ $ Q.E.D.
\end{pf}

\begin{thm}
  For the oscillator-qubit system, let $\mathscr{M}_{\rm inv}$ be a set of density matrices on the invariant manifold up to the second-order expansion and $\mathscr{D}(\mathscr{H}_B)$ be a set of density matrices on $\mathscr{H}_B$.
  Then, the partial trace map ${\rm tr}_A: \mathscr{M}_{\rm inv} \to \mathscr{D}(\mathscr{H}_B)$ is injective, but not surjective.
\end{thm}
\begin{pf}
  The injective property follows from the uniqueness of $\mathcal{K}^{G=0}$.
  On the other hand, the partial trace is not surjective because, from Corollary \ref{theorem:PureNegative}, any pure density matrix on $\mathscr{H}_B$ cannot be obtained by taking the partial trace of states in $\mathscr{M}_{\rm inv}$.
  $ \ $ Q.E.D.
\end{pf}

The positivity violation of $\mathcal{K}^{G=0}$ and thus the non-surjective property of the partial trace map can be understood from the correlation between the two sub-systems as discussed for composite isolated systems in \cite{Pechukas94}.
When $\epsilon = 0$, states in $\mathscr{M}_{\rm inv}$ are characterized by $\bar{\rho}_A \otimes \rho_B$ as mentioned above.
For these non-interacting separable states, any reduced state $\rho_B$ can be assigned to a valid total state.
When $\epsilon > 0$, on the other hand, states in $\mathscr{M}_{\rm inv}$ are no longer separable due to the interaction term.
At the second-order of $\epsilon$, they are characterized by entangled states.
Then, a pure reduced state cannot be assigned to any valid total state.

From this consideration, care should be taken if an initial state is determined in the reduced system.
In order to ensure that the corresponding total state is a valid one, the domain of initial reduced states must be restricted to ${\rm tr}_A (\mathscr{M}_{\rm inv})$, which is a subset of $\mathscr{D}(\mathscr{H}_B)$.
By linearity, extending the domain of $\mathcal{K}^{G=0}$ to $\mathscr{D}(\mathscr{H}_B)$ (all density matrices) is possible.
However, such extension leads to an initial total state that is not positive semidefinite and thus is unphysical.

\bibliography{root}             % bib file to produce the bibliography

\end{document}